\documentstyle{article}

\def\be{\begin{eqnarray}}
\def\ee{\end{eqnarray}}
\def\nn{\nonumber}

\begin{document}

\hfill hepth-0304178

\hfill{ITEP-TH-23/03}\\

\centerline{\Large{
Can Renormalization Group Flow End in a Big Mess?
}}

\bigskip

\vskip 1.0cm
\begin{center}
{\bf Alexei Morozov$^{*}$ and Antti J. Niemi$^{\dagger}$ } \\

\vskip 0.3cm

{\it $^{*}$Institute of Theoretical and Experimental Physics \\
B. Cheremushkinskaya, 25, Moscow 117259, Russia \\ \vskip 0.5cm
$^{\dagger}$Department of Theoretical Physics,
Uppsala University \\
P.O. Box 803, S-75108, Uppsala, Sweden }\\
\end{center}

\bigskip

\bigskip

\centerline{ABSTRACT}

\bigskip

The field theoretical renormalization group equations
have many common features with the equations of dynamical systems.
In particular, the manner how Callan-Symanzik equation
ensures the independence of a theory
from its subtraction point is reminiscent of self-similarity
in autonomous flows towards attractors.
Motivated by such analogies we propose that besides
isolated fixed points, the couplings in a renormalizable field
theory may also flow towards more general, even fractal
attractors. This could lead to Big Mess scenarios in applications
to multiphase systems, from spin-glasses and neural networks
to fundamental string (M?) theory. We consider various
general aspects of such chaotic flows. We argue that they
pose no obvious contradictions with the known properties of
effective actions, the existence of dissipative
Lyapunov functions, and even the strong version of
the $c$-theorem. We also explain the difficulties encountered
when constructing effective actions with chaotic renormalization
group flows and observe that they have many 
common virtues with realistic field theory effective actions. 
We conclude that if chaotic renormalization group flows are to be 
excluded, conceptually novel no-go theorems must be developed.

\noindent\vfill

\begin{flushleft}
\rule{5.1 in}{.007 in} \\
{\small  E-mail: \scriptsize MOROZOV@ITEP.RU, $~~~$
NIEMI@TEORFYS.UU.SE}
\end{flushleft}
\vfill\eject
\noindent
{\bf \Large 1. Introduction}
\vskip 1.0cm

{\baselineskip 0.6cm

The concept of a renormalization group (RG) flow is a basic notion
both in quantum field theory and string theory
\cite{RGgen}-\cite{MM}.
In the Wilsonian approach, the RG equation
describes a flow in the space of operators towards a subspace
of relevant and marginal operators. This subspace
can be viewed as a functional attractor for the flow.
The behaviour within the functional attractor is then
determined by the $\beta$-functions for the relevant couplings.
Conventionally one expects that these couplings flow towards
attractors which consist of a finite number of
isolated fixed points. This ensures that
the limit of the RG flow yields a definite quantum 
field theory with destined values for its couplings.

However, already in \cite{Wor} Wilson noted that the RG
flow of the couplings could approach attractors which
are more elaborate than plain isolated fixed points. 
In particular, he suggested that
the end of a RG coupling flow could
be a limit cycle. Some evidence in support of this conjecture
has been recently provided in \cite{cycles1}-\cite{cycles3},
where a coupling flow with a (periodic) blow-up is interpreted
in terms of a limit cycle. 

Here we
shall inquire whether a RG flow could indeed tend towards
a nontrivial attractor, with even a {\it fractal} structure
corresponding to a chaotic flow in the
space of couplings. Such chaotic flows could
lead to interesting Big Mess scenarios in various applications
of quantum field and string theories. Indeed, we suggest
that the commonly accepted dogma that
RG flows can only approach plain, discreet fixed
points is in an apparent contradiction with
the existence of multiphase systems described by spin glasses
and neural networks, which are expected to
emerge as the IR endpoints of RG flows from simple microscopic
Hamiltonians. Furthermore, it should be only natural to
speculate that the IR finality of the full (but yet
to be discovered) string theory possesses complex
multiphase structures \cite{M2} with a variety
of quantum field theories, strings and brane models,
emerging at the ends of some as yet unidentified
chaotic RG evolutions. These flows are expected originate
from a relatively simple (fundamental microscopic) system,
for example from the celebrated $M$-theory \cite{Mth1,Mth2}.
We propose that such chaotic RG scenarios are realizable, 
very much like condensed matter physics with its highly complex
long-distance structures emerges
from a simple microscopic Hamiltonian. Indeed, if complex systems are to
appear as asymptotic IR limits of some RG flows it should
be clear that these flows can not be towards a finite set of
plain isolated fixed points but rather towards more general,
even {\it strange} attractors with self-similar structures over
several orders of magnitude. 

We emphasize that we are not
considering the possibility of chaos in an underlying field
theory. For example, in the classical Yang-Mills theory chaotic
behaviour has already been well established \cite{YMch}.
Consequently such chaotic behaviour will not be considered
here. Obviously, a chaotic RG flow also necessitates the consideration
of field (string) theories with at least three couplings.

In the present article we shall be interested in the possibility
of chaotic RG flows
in the IR limits of quantum field and string
theories. While we are not in a position to present concrete
examples of theories which exhibit a chaotic RG flow,
we do have a number of plausibility arguments which
support their existence. Furthermore, we can understand and
explain the difficulties encountered in their
constructions. At least, our work should
motivate the derivation of no-go theorems.
But at the moment we do not see any
immediate contradictions between the existence of
chaotic RG flows and known properties of field and string
theories. Indeed, we believe that
the potential existence of a chaotic RG flow is an
important question, and either examples should be searched
for or then
conceptually novel no-go theorems forbidding
chaotic RG behaviour should be established.

Exact RG group equations are operative everywhere in the
space ${\cal M}$ of couplings $\{t^i\}$. Occasionally, and
in particular in the context of string theories, these
couplings are identified with the moduli space of
the theory under consideration. The Wilsonian approach to
the exact RG is based on an effective action,
\[
{\cal F}_{\cal A}(t|\varphi) \sim \log Z_{\cal A}(t|\varphi),
\]
\be
Z_{\cal A}(t|\varphi) = \int_{\cal A} D\phi
\exp \left( \sum_{i} t^i{\cal O}_i(\phi)\right) \ \
\ \ \ \ ({\rm with} \ \
\left.\phi\right|_{\partial{\cal A}} = \varphi)
\label{ea}
\ee
The integration extends over a functional
space ${\cal A}$ of fields $\phi$. The effective action
${\cal F}$ depends both on the background fields $\varphi$ and on
a functional form which is parameterized
by the couplings $t^i$. Consequently it is
a section of a line bundle over
background fields $\varphi$ and the manifold ${\cal M}$
of couplings $t^i$. The Wilsonian RG flow (in the sequel
we consider mainly flows from the UV to the IR) describes
the change of ${\cal F}$ under the change of
${\cal A}$, when some of the background fields are integrated
out. In the simplest case these background fields
are Fourier components with momenta exceeding
some normalization scale $\mu^2$. The boundary
$\partial {\cal A}$ can have a generic shape. It
is parametrized by Whitham times $T_\alpha$,
and the Wilsonian RG equations can be understood as
evolution equations in all possible
Whitham directions \cite{Whith,GMMM}.
This leads to a relation between the RG flow and
the concept of self-similarity
(or functional similarity in the terminology of \cite{KS})
between effective actions evaluated at different
Whitham times. The present version of renormalizability
then makes exact RG flows a part of the general
theory of dynamical systems \cite{dys}.

The specific
feature of a RG flow when viewed as a dynamical system
is that it involves the effective
action. As a section in a line bundle the effective action
is in general a multivalued function
of the background fields and
couplings. Consequently there is an element of local
integrability, that may be eventually lost due to global 
obstructions. Furthermore, since the effective
action is defined by a (functional) integral it
possesses extensive symmetries relating to changes of
the integration variables (quantum fields) $\phi$. These
symmetries are usually expressed in terms of Ward
identities (or Picard-Fucks equations) \cite{FaSl}-\cite{CDSW}.
For an exact RG, the number of independent couplings $t^i$
must also be large - in fact infinite - to ensure that the
operators involved indeed form a complete basis of functionals.
Furthermore, when we ignore the background fields $\varphi$, the
exact RG equations \cite{Pol} for the effective action
acquire a Callan-Symanzik form, which is linear in
the derivatives {\it w.r.t.} couplings,
\be
\dot Z(t) =
\frac{\partial Z(t)}{\partial s} =
\sum_i \beta^i(t)\frac{\partial Z(t)}{\partial t^i}, \nn \\
\dot{\cal F}(t) =
\sum_i \beta^i(t)\frac{\partial {\cal F}(t)}{\partial t^i}, \nn \\
\beta^i(t) = \dot t^i = - \mu \frac{\partial t^i}{\partial \mu}, \ \
\ \ (s = -\log\mu)
\label{baRG}
\ee
This functional form emerges for a {\it strongly}
complete basis of 
operators ${\cal O}_i(\phi)$ (for definitions see \cite{MM})
which includes all {\it linearly} independent generators.
(For a {\it weakly} complete basis which includes
all {\it algebraically} independent operators, the RG
equations contain higher derivatives of $Z(t)$ \cite{MM}.)

\vskip 0.5cm

In the next section we shall first consider certain general aspects
of the coupling flow. We then review some of the assumptions
that underlie RG flows, i particular the $c$-theorem(s). In section
3 we first relate a few field theory $\beta$-functions to the $c$-theorems.
We then argue by considering the Lorenz model, that the properties
of these coupling flows are not inconsistent with chaotic RG flows.
In section 4 we consider limit cycles from the point of view
of RG flows, and inspect vorticity as a RG scheme independent tool
for describing multicoupling flows. In section 5 we study  
model effective actions as toy models for reproducing realistic
scaling properties of field theories. In section 6 we explain
how to construct model effective actions from the $\beta$-function
flows. In particular, we explain how the construction fails in
case of chaotic flows and suggests this parallels the problems
encountered in constructing actual field theory effective actions.
This also explains why it is very hard to construct actual field theory
models with chaotic RG flow.
In section 7 we consider possible extensions and scenarios for
realizing chaotic RG flows, including spectral flow in general and
in particular in stringy context. We conclude with some suggestions
on Early Universe models. 

\vfill\eject

\noindent
{\bf \Large 2. The $c$-theorems}
\vskip 1.0cm
The idea of a chaotic attractor is actually not
too distant from a quantum field theoretical RG
equation. This can be seen already
by inspecting the functional form
of familiar one-loop single coupling $\beta$-functions.
For example, in $d=2+\epsilon$ dimensions
the one-loop $\beta$ function of the $O(N)$
nonlinear $\sigma$-model has the functional form
\be
\dot g \ = \ \epsilon ag(1 - g)
\label{vert}
\ee
which coincides with the form of the Verhulst equation
of population growth \cite{stroga}. In its discretized
version, this clearly relates to the logistic equation
\be
x_{n+1} = c x_n (1 - x_n)
\label{logis}
\ee
which is the classic example of an iterative
equation with chaos.

But for chaos in a continuous RG flow, we need at least
three independent couplings: In a renormalizable field
theory with several couplings $t^i$,
the RG flow is described by the following autonomous linear
system (recall that when considering flows towards the IR limit
we have $s = - \ln \mu $ which means that the ensuing
(IR) $\beta$-functions are positive in asymptotically
free models)
\be
- \mu \frac{\partial t^i}{\partial \mu} \ = \
\frac{\partial t^i}{\partial s} = \beta^i(t)
\label{dynsys}
\ee
Conventionally, one assumes that in a field theory
this IR flow is asymptotically approaching an isolated
fixed point which is hyperbolic. For classification purposes
we may then evoke the Hartman-Grobman theorem \cite{HGt} which
allows us to consider $\beta$-functions which are
linear in the couplings,
\be
\beta^i(t) \ \approx \ {B^i}_j t^j \ + \ {\cal O}(t^2)
\label{hart}
\ee
(${B^i}_j$ constant). We note that this may not be attainable
by conventional changes in the renormalization scheme, which are
analytic diffeomorphisms on the space of couplings of the form
\be
t^i \ \to \ \ \tilde t^i (t) \ = \ {A^i}_j t^j \ + \
{\cal O}(t^2)
\label{diffeo}
\ee
where ${A^i}_j$ is a constant nonsingular matrix with
positive eigenvalues.
Obviously, the first term in (\ref{hart})
describes only integrable, linear flow along the eigendirections
of the matrix ${B^i}_j$. But it is widely respected
that even in the presence of apparently very simple
non-linear corrections, such as in the case of the R\"ossler
system \cite{Rossler}\footnote{Notice that this
representation of the R\"ossler equations
differs from the conventional one \cite{Rossler} where
the last equation reads $\dot z = \tilde b + z(x + \tilde c)$,
by a linear transformation of variables:
$ x \rightarrow x + ab, \ \ y \rightarrow y - b, \ \ 
z \rightarrow z + b $ so that  $\tilde b = -bc$ and $\tilde c = c -ab$.
\label{ross}}
\be
\dot x \ = \ - y - z \nn \\
\dot y \ = \ x + a y \nn \\
\dot z = bx + cz + x z
\label{rossler}
\ee
the equations (\ref{dynsys}) can possess a multitude of asymptotic
behaviours as the flow-time $s$
goes to infinity. Besides isolated fixed points
and limit cycles, the trajectories can also approach
chaotic {\it strange} attractors with complex geometries and
fractal dimensions.

Notice that the equation (\ref{dynsys}) is renormalization scheme
covariant, {\it i.e.} form invariant under the diffeomorphisms
(\ref{diffeo}). Indeed, under a {\it general} coordinate transformation
$t^i \rightarrow \tilde t^i(t)$ the $\beta^i$ transform as
components of a vector field
$$
\tilde\beta^i(\tilde t) = \frac{\partial \tilde t^i}{\partial t^j}
\beta^j(t),
$$
Infinitesimally, for $\tilde t^i = t^i + \epsilon^i(t)$
we then have
$$
\delta\beta^i(t) = \tilde\beta^i(t) - \beta^i(t) =
\beta^j\frac{\partial\epsilon^i}{\partial t^j} -
\epsilon^j\frac{\partial\beta^i}{\partial t^j} =
-\left({\cal L}_\epsilon\beta\right)^i
$$
which is the Lie derivative of $\beta$ along $\epsilon$.
Consequently we can interpret the flow (\ref{dynsys})
geometrically in a renormalization scheme independent
manner, with $\beta^i$ a vector field
in the tangent bundle of ${\cal M}$.

In formal quantum field theory investigations one
traditionally assumes that the flow (\ref{dynsys})
can only tend towards isolated fixed points. For this,
the vector field $\beta^i$ is subjected
to a variety of conditions. A pivotal requirement is that
the renormalization group flow must be irreversible.
According to Zamolodchikov's ({\it weak}) $c$-theorem
\cite{Zam,cthm} this irreversibility is ensured by the
existence of a Liapunov function $c(t)$ which is
monotonically decreasing along the RG flow towards
the IR
\be
\frac{d c(t)}{ds} =
\beta^i(t)\frac{\partial c(t)}{\partial t^i}
< 0
\label{weakc}
\ee
Occasionally one also assumes that the Liapunov
function is positive semidefinite $c(t) \geq 0$.
It may then be related to the number of degrees
of freedom in the theory.

A {\it strong} version of the $c$-theorem states that
the vector field $\beta^i$ is a gradient,
\be
\beta^i(t) =
- G^{ij}(t)\frac{\partial \sigma(t)}{\partial t^j}
\label{gra}
\ee
with a symmetric metric $G_{ij}(t) = G_{ji}(t)$.
Furthermore, if $G_{ij}(t)$ is positive-definite
as is usually assumed in the strong $c$-theorem,
the generating function $\sigma(t)$ is a Lyapunov function:
\be
\dot \sigma =
\frac{d\sigma}{ds} = -\sum_{ij} G^{ij}
\frac{\partial \sigma}{\partial t^i}
\frac{\partial \sigma}{\partial t^j}
< 0
\label{morse}
\ee
Furthermore, in the case of two-dimensional field
theories it can be argued that $\sigma(t)$ is a (peferct)
Morse function \cite{morse} with its (isolated) critical points
corresponding to conformal field theories. In that case
(\ref{morse}) becomes a gradient flow between the critical
points of $c$, {\it i.e.} between different
conformal field theories.

We note that these statements on the RG flow are
renormalization scheme independent, and intrinsically
geometric.

\vfill\eject

\noindent
{\bf \Large 3. Some Examples}
\vskip 1.0cm

We shall now argue, with examples, that many of the
widely accepted properties of the RG flow do not exclude
strange attractors from appearing in the IR limit of the
flow. We start by considering examples of field theoretical
RG coupling flow equations from the perspective
of the $c$-theorems.

An important example of a RG coupling flow
with three independent couplings (the minimal
number required for a chaotic flow) is the
$U(1)\times SU(2) \times SU(3)$ standard model in
four dimensions. At the two-loop level, and ignoring
the contribution from the Higgs sector, the three gauge
couplings $t^i$ flow according to \cite{macha}
\be
\dot t^i \ = \ g^{ij} \partial_j F
\label{stand1}
\ee
where the metric $g^{ij}$ has the form
\be
g^{ij} \ = \ - \delta^{ij}(b^i + \sum_k {b^i}_k (t^k)^2)
\label{stand2}
\ee
with
\be
16 \pi^2 b^1 \ = \  \frac{4}{3} n + \frac{1}{10} \nn \\
16\pi^2 b^2 \ = \ \frac{22}{3} - \frac{4}{3} n - \frac{1}{6} \nn \\
16\pi^2 b^3 \ = \ 11 - \frac{4}{3}n
\nn
\ee
\be
{b^i}_k \ = \ \frac{1}{(16\pi^2)^2}
\left(\begin{array}{ccc}
0 & 0 & 0 \\
0 & \frac{136}{3} & 0 \\
0 & 0 & 102
\end{array}\right) \ - \ \frac{n}{(16\pi^2)^2}
\left(\begin{array}{ccc}
- \frac{19}{15} & - \frac{1}{5} &
- \frac{11}{30} \\
\frac{3}{5} & \frac{49}{3} & \frac{3}{2} \\
\frac{44}{15} & 4 & \frac{76}{3}
\end{array}\right)
\ee
where $n$ is the number of generations and $F$ is a (degenerate)
Morse function for the critical point at $t^i = 0$,
\be
F(t) \ = \ \frac{1}{4}\left[(t^1)^4 - (t^2)^4 - (t^3)^4 \right]
\ee
Here the indefiniteness of $F(t)$ (which is usually excluded
by $c$-theorems due to its {\it naive} contradiction
with unitarity) reflects UV asymptotic freedom
of the non-abelian components in the model.

The first ($b^i$) term in (\ref{stand2}) is the one-loop contribution,
the second (${b^i}_k$) term is the two-loop correction.
Note that the entire two-loop contribution can be viewed
as a correction to the one-loop metric. We also note that
depending on $n$, the metric can have different signatures.
Furthermore, depending on $t^i$ the signature of the
two-loop metric can differ from the signature of the
one-loop metric. While these observations
on the qualitative attributes of the
metric {\it as such} can hardly have much relevance to the Physics
of Standard Model they are still instructive in
revealing the variety of properties that
coupling flows in quantum field theories share.

As another example, where already at the one-loop level
the signature of the metric depends on the relative (small)
values of the couplings, we consider the standard
Yukawa coupling between a pseudoscalar meson and a Dirac fermion.
There are now two couplings, the Yukawa coupling $g$
and the quartic self-coupling $\lambda$ of the pseudoscalar.
At the one-loop level the flow equations are (in $d=4+\epsilon$
dimensions, with $(2\pi)^d N_d$ the area of the
unit sphere in $d$ dimensions)
\cite{2loopYukawa}
\be
\dot g \ = \ \frac{\epsilon}{2} g + \frac{5}{2} N_d g^3 + \ ...\nn \\
\dot \lambda \ = \ \epsilon \lambda + \frac{3}{2}
N_d \left( \lambda^2 + \frac{8}{3} \lambda g^2 - 16 g^4\right) + \ ...
\ee

The origin $g = \lambda = 0$ is a critical point with
its stability depending on the sign of $\epsilon = d-4$.
If we introduce the non-degenerate Morse function
\be
F = \frac{1}{2}(g^2 + \lambda^2)
\ee
these equations can be written in the gradient-flow form
\be
\dot g \ = \ G^{gg} \partial_g F \ = \ \frac{1}{2}(\epsilon + 5 N_d g^2)
\partial_g F
\nn \\
\dot \lambda \ = \ G^{\lambda \lambda} \partial_\lambda F \ = \
( \epsilon + 4 N_d g^2 - 24 N_d \frac{g^4}{\lambda} + \frac{3}{2}
N_d \lambda) \partial_\lambda F
\label{sqed}
\ee
Obviously the metric can be either positive definite,
negative definite or indefinite depending on the
parameters, and the relative strength of the couplings.
To some extent this can be compensated, by adjusting the
relative signs of the two terms in $F$. But since
the metric can also change its signature at non-vanishing,
even small values of $g$ and $\lambda$ it can not be made positive
everywhere. This reflects the fixed point structure
of the theory on the $(g,\lambda)$ plane.

As a third example we consider the model which has
been studied in \cite{cycles2} as a candidate for limit
cycle behaviour in the coupling flow.
This is the $su(2)$ level $k=1$ Wess-Zumino model,
at the one-loop level its couplings $g$ and $h$ flow according
to
\be
\dot g \ = \ - h^2 \ = \ - h \partial_g (hg)
\nn \\
\dot h \ = \ - g h \ = \ - h \partial_h (hg)
\label{wzw}
\ee
Consequently in these coordinates the RG equations have the gradient
flow form but the $c$-function is not a non-degenerate Morse
function. Furthermore, unless $h>0$ the metric is
not positive definite. In the present model the exact
$\beta$-functions are also
known. The ensuing RG flow equations are \cite{lukia},
\cite{cycles2}
\be
\dot g \ = \ - \frac{h}{2-g} \partial_g (hg) \nn \\
\dot h \ = \ - \frac{h}{2-g} \partial_h (hg)
\label{luki}
\ee
In these (isothermal) coordinates
we then have a metric tensor which is singular, and
a $c$-function which is not
a nondegenerate Morse function. Notice in particular
that the $c$-function
does not receive corrections beyond the one-loop, all higher order
corrections lead only to a modification of the metric.
We note that the ensuing coupling flow on the $(g,h)$
plane has a blow-up at
finite value of the flow parameter $s$. It has been suggested
\cite{cycles1}, \cite{cycles2} that this could
be interpreted as a flow towards a limit cycle (see below).

From these examples it is clear that the $\beta$-functions
that appear in quantum field theories are not
always of the form suggested by the various versions of the $c$-theorem.
In particular, while the equations do admit the gradient flow form
(\ref{gra}) with (trivially) symmetric metric,
the $c$-functions are not necessarily nondegenerate Morse functions nor
are the metric tensors necessarily positive definite or even regular
everywhere in the space of couplings. While in some 
models such deviations from the
$c$-theorems could be attributed to limitations of perturbation theory
and removed by higher order corrections, the
example (\ref{luki}) shows that this is not always the case.
Moreover, even though the flows admit a gradient representation,
they do not in general describe simple, structureless laminar flows towards
isolated fixed points along gradients of the $c$-functions:
If we consider the vector fields
which appear on the {\it r.h.s.} of the flow equations, in all of
the three examples, we note that each of the vector fields
carries a non-trivial {\it vorticity} two-form (defined
as the exterior differential of the covector of the
$\beta$-functions)
\be
\omega_{ij} \ = \ \partial_i \beta_j - \partial_j \beta_i
\label{vorti}
\ee
Curiously, for the standard model the vorticity vanishes at
the one-loop level and appears only at the two-loop level (and beyond).
This suggests that higher loop corrections have a qualitative effect
on the theory. Both in the pseudoscalar model and WZW model vorticity
is present and regular already at the one-loop level. But
from (\ref{luki}) we find that the vorticity in the full theory
can have a singularity at finite coupling
\be
\omega \ = \ \epsilon_{ij} \partial_i \beta_j \ = \ \frac{g(g-2)
- h^2}{(g-2)^2}
\nn
\ee
reflecting the blow-up at finite flow time.
We note that since vorticity is a closed two-form it
can be made constant in a neighborhood of any regular
point by diffeomorphisms. But if it is non-vanishing in one
coordinate system it remains non-vaninshing in all coordinate
systems. Consequently vorticity is a renormalization scheme
independent characteristic of the flow.

Maybe somewhat surprisingly several qualitative aspects of our
examples, as well as many general assumptions in the $c$-theorems,
can also be realized in chaotic systems. As an example,
we consider the three dimensional Lorenz system \cite{Lor},
\be
\dot x = -\sigma x + \sigma y \ = \ \beta^x \nn \\
\dot y = rx - y - xz \ = \ \beta^y \nn \\
\dot z = xy - b z \ = \ \beta^z
\label{Lore}
\ee
($\sigma, r, b$ are positive constants). We start
by introducing a (Liapunov) $c$-function,
an arbitrary positive semidefinite function
$\rho \geq 0$ in $R^3$ which we advect along the Lorenz
flow. This is described by the conservation of the
current
\[
j^\mu \ = \ (\rho, \rho \beta^i)
\]
\[
\frac{\partial \rho}{\partial t} + 
\nabla \cdot (\rho {\vec \beta}) \ = \ 0
\]
which reduces to
\be
\frac{d\rho}{dt} \ = \ \frac{\partial \rho}{\partial t}
+ {\vec \beta} \cdot \nabla \rho \ = \ - (\nabla \cdot {\vec \beta}) \rho
\ = \ -(\sigma + b + 1 ) \rho \ \leq \ 0
\label{lorirr}
\ee
Consequently {\it any} positive function on $R^3$ which is advected
by the Lorenz flow satisfies the irreversibility requirement
of the $c$-theorem.

Consistent with the $c$-theorems, the Lorenz system
can also be presented in the
form of a gradient flow (\ref{gra}) with a symmetric metric. 
For example, if
we introduce a non-degenerate Morse function for the
critical point at the origin
\be
F(x,y,z) \ = \ x^2 + \sigma(y^2 + z^2)
\label{morlor}
\ee
we can write the Lorenz system as
\be
\dot x \ = \ g^{11} \partial_x F \ = \ \sigma \frac{y-x}{2x} \partial_x F
\nn \\
\dot y \ = \ g^{22} \partial_y F \ = \ \frac{rx - y - xz}{2\sigma y}
\partial_y F \nn \\
\dot z \ = \ g^{33} \partial_z F \ = \ \frac{xy - bz}{2\sigma z}
\partial_z F
\ee
As in the previous three field theory examples, the positive definiteness
of the metric tensor $g_{ab}$ depends on the relative values
of $x,y$ and $z$. Like the metric in (\ref{luki}),
this metric is also non-singular except for the critical points
of the flow (and the $x,y,z=0$ lines)
while $F$ is both a Morse function for the critical point
at origin, and a global Liapunov function when $r<1$ since
\be
\frac{dF}{dt} \ = \ -2\sigma( \left[x -
\frac{1}{2}(1+r)y\right]^2 + \frac{1}{4}
(1-r)(3+r) y^2 + 2 \sigma b z^2 ) \ < \ 0 \ \ \ \ \ \ (x,y,z \not=0)
\nn
\ee
Consequently, at this level of analysis we do not see much
difference in the qualitative properties of the (chaotic)
Lorenz model and the coupling flows in our three field theory
examples. In particular, nothing appears to prevent chaos 
from occuring in RG flows as well.

\vfill\eject

\noindent
{\bf \Large 4. Limit Cycles and Vorticity}
\vskip 1.0cm

The existence of a positive definite metric with a $c$-function
which decreases along the orbits can also be satisfied
by flows without simple fixed points. For example, if we take
$$
c(t,\bar t) = (t\bar t - a^2)^2
$$
the entire cycle $t = ae^{i\theta}$ forms an attractor.
In order to induce a motion along this cycle
we then consider ($\eta > 0$)
\be
\dot x = - \eta y \ = \ -\beta^x \nn \\
\dot y = \eta x \ = \ -\beta^y
\label{rot}
\ee
This flow is consistent with the
strong $c$-theorem,
with a positive-definite metric
\[
G^{ij} = \eta \delta^{ij}(x^2+y^2)
\]
and a (multivalued) Liapunov $c$-function which is
monotonically decreasing along the flow,
\[
c(x,y) = \arctan (y/x)
\]
The ensuing flow has constant vorticity,
\[
\omega \ = \ \epsilon_{ij}\partial_i \beta_j \ = \ 2 \eta
\]
Such flows may actually be realistic in models where the
coupling (moduli) space has nontrivial topology,
with non-vanishing $\pi_1({\cal M})$
({\it cf.} \cite{Nov}).
Indeed, since the $c$-function
is highly nonlinear the flow could be viewed as
a non-perturbative one. We are not aware
of any apparent reason why this kind of
flow should in general be excluded by the $c$-theorem.

There is a simple generalization of (\ref{rot}) to non-constant
vorticity,
\be
\dot x = -y^k,\nn \\
\dot y = x^k
\label{odd}
\ee
with $k$ an integer. This flow is of the form of strong $c$-theorem,
with
$$
c(x,y) = y^{1-k} - x^{1-k}
$$
The corresponding metric tensor is
$$
g^{ij} = \frac{\delta^{ij}}{k-1} (xy)^{k}
$$
For $k = 2n$ this metric is positive on the $(x,y)$ plane, 
and the $c$-function has the form of
a (degenerate) Morse function as expected by the strong $c$-theorem.
But for $k = 2n+1$ {\it i.e.} odd,
the metric is in general not positive.
It is instructive to consider in more detail the reasons for
the failure of strong $c$-theorem when $k$ odd.
For this we note that the flow possesses
a conserved quantity, $x^{k+1} + y^{k+1} = const$. This suggests us
to change variables
\be
x = R\cos^{1/(n+1)} \theta \nn \\
y = R\sin^{1/(n+1)} \theta
\label{cosfrac}
\ee
so that
\be
\dot R = 0, \nn \\
\dot \theta = \frac{\partial c(\theta)}{\partial \theta}
\ee
where $c(\theta)$ is $R$-independent. Consequently, in these coordinates
it appears that the flow is consistent with
the strong version of the $c$-theorem
with a positive definite metric
\[
dR^2 + R^2d\theta^2
\]
But when we transform back to the cartesian $x,y$
coordinates,
\be
\left(
R^2\left(\frac{\partial X}{\partial R}\right)^2
+ \left(\frac{\partial X}{\partial \theta}\right)^2\right)
dX^2 +
2\left(
R^2 \frac{\partial X}{\partial R}
\frac{\partial Y}{\partial R}
+ \frac{\partial X}{\partial \theta}
\frac{\partial Y}{\partial \theta}
\right) + \nn \\
+ \left(
R^2\left(\frac{\partial Y}{\partial R}\right)^2
+ \left(\frac{\partial Y}{\partial \theta}\right)^2\right)
dY^2
\ee
We find that due to the fractional powers of trigonometric
functions in (\ref{cosfrac}) the ensuing metric
is not positive definite, for example when $k=1$ we have
\be
G^{XX} = \left(\frac{R\sin\theta}{2\sqrt{\cos\theta}}\right)^2
+ \left(\frac{R\sqrt{\cos\theta}}{2}\right)^2 =
\frac{R^2}{4\cos\theta}
\ee
which is not positive definite.

Notice that besides a gradient flow,
the flows (\ref{rot}), (\ref{odd}) can also
be interpreted in terms of a symplectic (Hamiltonian) flow
of the form
\be
\beta^i(t) = \sum_j \omega^{ij}
\frac{\partial H }{\partial t^j}
\label{Ham}
\ee
with a closed symplectic two-form
$\omega = \omega_{ij}dt^i\wedge dt^j$
and a Hamiltonian $H$. In particular,
(\ref{rot}) is the harmonic oscillator.

We note that the presence of such limit cycles in RG
flows is quite important from the point of
view of (possible) chaotic flow. Obviously,
a limit cycle behaviour is much easier to analyse
than a chaotic behaviour. But in addition,
limit cycles can also provide a period doubling (Feigenbaum)
route to chaos (for flows with at least three couplings)
\cite{Feig}.
Indeed, if for some value of external control
parameters a system exhibits a stable limit cycle,
its stability can be lost by period doubling
when the control parameter changes. When this happens,
the attractive cycle becomes repelling and
instead there is a new attractive limit cycle
which exhibits period doubling and links around
the previously stable cycle. When the control parameter is
varied further, there will be additional period
doublings and eventually these bifurcate
into an infinitely long limit cycle, with a
fractal structure. The R\"ossler system
(\ref{rossler}) is the simplest three dimensional
example which exhibits this Feigenbaum route to
chaos by consequtive period-doublings in its limit cycles,
while in the Lorenz system (\ref{Lore}) the transition
to chaos is by intermittency \cite{stroga}.

Unfortunately, it appears to be quite difficult to
find RG limit cycles. While periodic dependency
on the coupling constants in the $\beta$-functions
has been well established, for example in the context
of topological charge renormalization \cite{TC}, this
is not sufficient for obtaining a limit cycle.
Until now, only one example of limit cycle behaviour
in RG equations has been constructed
\cite{cycles1}-\cite{cycles3},
but unfortunately this example can not be considered
fully satisfactory. Essentially, the flow is of the form
\be
\dot g = h^2 + g^2,
\nn \\
\dot h = 0,
\nn
\ee
which exhibits a {\it blow-up}, rather than smooth {\it cyclic}
behaviour. We note that for small values of $g$ these equations
are quite similar to the conventional asymptotically free RG equations
(with $h=0$), the only difference is that whenever the second
coupling constant $h\neq 0$ the flow becomes accelerated.
The ensuing flow
\[
g(\mu) = h\cot (h\log\mu)
\]
is formally periodic in the sense that
\[
g(e^{\pi/h}\mu) = g(\mu)
\]
but a discontinuous {\it jump} from $g = +\infty$ to $g=-\infty$
is necessary in order to close the cycle. Unfortunately,
concrete examples with continuous limit cycle behaviour in realistic
field theory models have not yet been found.

The previous examples underline the importance of developing general
classification schemes for $\beta$-function
flows in multidimensional cases. As we have proposed, 
a local approach
could be based on the Hartman-Grobman theorem
\cite{HGt} on hyperbolic fixed points. This suggests
that for classification purposes
we consider flow equations with $\beta$-functions which
are at most bilinear in the couplings, the bilinearities representing
either corrections to, or deviations from hyperbolic behaviour.
This subclass
of flows is particularly interesting in three dimensions,
since bilinear nonlinearities are sufficient for a chaotic
flow to occur. For this we consider three dimensional
flows of the form
\be
\dot x^i \ = \ \beta^i \ \ \ \ \ \ \ (i=1,2,3)
\label{3dflow}
\ee
The three dimensional (co)vector $\beta_i$ can be presented
using a Glebsch decomposition
\[
\beta _i \ = \ \mu \partial_i \rho + \partial_i \gamma
\]
with three functions $\mu, \rho, \gamma$. When $\mu = 0$
the vorticity 
\[
\omega_i \ = \ \epsilon_{ijk} \partial_j \beta_k \ = \
\epsilon_{ijk} \partial_j \mu \partial_k \rho 
\] 
vanishes,
limit cycles are absent and the ensuing advection of the couplings is
laminar, non-chaotic gradient flow.
But whenever $\mu$ is non-vanishing
the vorticity is non-vanishing and either constant
or linear in the couplings. We now argue that
both for constant and linear vorticity, the advection
can be chaotic. (For a flow with vanishing vorticity,
chaotic advection is hardly possible.) Recall that since
vorticity is a closed two-form, we can introduce a
diffeomorphism which maps it into a constant in a
neighborhood of a regular point.
But it can not be made to vanish in that neighborhood
by diffeomorphisms.

Consider first the Lorenz equations (\ref{Lore}).
The $\beta$-functions can be Glebsch-decomposed according
to
\[
\beta_i \ = \ xy^2 \partial_i\left( \frac{z+\sigma - r}{y}
\right) + \partial_i \left( \sigma xy - \frac{1}{2}(
\sigma x^2 + y^2 + bz^2) \right)
\]
and for the vorticity we find
\be
\omega = -z dx\wedge dy + (2x-\sigma) dy\wedge dz - y dz\wedge dx
\label{lormon}
\ee
Notice that on the surface $x^2+y^2+z^2 = const$
this involves a term which represents $H^2(S^2)$,
the (unique) volume element of $S^2$. Consequently
the vorticity (\ref{lormon}) of the Lorenz system is a
representative of the monopole bundle in $R^3$.

For the R\"ossler equation (\ref{rossler}) we find the
vorticity (here we use the original form of these equations,
see footnote (\ref{ross}) in connection of equation (\ref{rossler}).)
\be
\omega \ = \ -2 dx \wedge dy + (1+z) dx\wedge dz \ = \
- 2 dx \wedge dy + \frac{1}{2}dx \wedge d(1+z)^2
\ee
Consequently the vorticity can be made
constant by
a quadratic diffeomorphism, but at the expense of loosing
the quadratic nature of the equations.

A chaotic system with simultaneously quadratic nonlinearities and
constant vorticity can also be constructed, for example by
combining the behaviour of the $\beta$-functions in (\ref{wzw})
and (\ref{rot}) into the following three dimensional flow
\be
\beta_x \ = \ \frac{1}{2} z + y - 1 + px + \frac{1}{2} y^2 \nn \\
\beta_y \ = \ - x + xy - y -1 \nn \\
\beta_z \ = \ - \frac{1}{2} x + a z
\label{cc}
\ee
This can be shown to be chaotic for appropriate values of
the parameters $p,a$ \cite{omaun}. Vorticity is linear
\[
\omega \ = \ -2 dx \wedge dy + dz \wedge dx
\]

In $D$ dimensions we can introduce a generalization
of the Glebsch decomposition
\[
\beta_i \ = \ \sum_{k=1}^{N} \mu_k \partial_i \rho_k
\ + \partial_i \gamma
\]
Where $2N=D$ for $D$ even, and $2N=D-1$ for $D$ odd, and $\gamma$
is present only for odd $D$. Presumably a necessary condition
for chaotic advection in $D \geq 3$ is that the ensuing vorticity
\[
\omega_{ij} \ = \ \sum_{k=1}^{2N} (\partial_i
\mu_k \partial_j \rho_k - \partial_j \mu_k \partial_i \rho_k)
\]
is non-vanishing. Since a non-trivial vorticity appears to be
generic for $\beta$-functions in quantum field theories, additional
restrictions are then needed to exclude chaotic advection of the
couplings.

\vfill\eject
\noindent
{\bf \Large 5. RG Flows And Model Effective Actions}
\vskip 1.0cm

It appears that nontrivial vorticity is generic for
the coupling flows, to the extent that even the one-coupling
$\beta$-functions can be related to flows with vorticity.
For this, we consider the familiar one-loop formula
\be
\frac{\partial}{\partial s}
{\left(\frac{1}{g^2}\right)} = -2b,
\label{olb}
\ee
which appears {\it e.g.} at one-loop four dimensional
Yang-Mills theories, {\it cf. } (\ref{stand1}), (\ref{stand2}).
We then introduce a Coleman-Weinberg type model
effective action (for {\it strong} fields). This
is an ordinary function of a single real variable $F$,
\be
{\cal F}(g,h|F) = \frac{1}{g^2} F^2 + h F^2\log F,
\label{toyF}
\ee
It turns out that many properties of RG flows can be
understood by inspecting the scaling properties of
such ordinary functions. Indeed, since
the $\beta$-functions in (\ref{dynsys}) are independent
of spacetime coordinates, we can expect that the
essential aspects of RG flows are independent of spacetime
coordinates and can be studied in tems of such ordinary functions
({\it i.e.} model effective actions) in lieu of the actual 
quantum effective actions. 
This is certainly the case in theories where we can have an
effective potential, such as Higgs models.

When we introduce the scaling $F \rightarrow \lambda^2 F$ and
${\cal F} \rightarrow \lambda^4 {\cal F}$ in (\ref{toyF}),
we get
\be
\frac{\partial}{\partial s}{\left(\frac{1}{g^2}\right)} =
-\frac{2\dot g}{g^3} = h, \nn \\
\dot h = 0
\label{colwei}
\ee
The solution
\be
h = const = -2b \nn \\
g^{-2} = 2b\log\mu \nn
\ee
of these equations
leads to the $\beta$-function
(\ref{olb}). Nevertheless, the system (\ref{colwei})
has a non-trivial vorticity
\be
\omega \ = \ dh \wedge dg^{-2} \ \not= \ 0
\ee
More generally, if we consider model effective
actions of the form
\be
{\cal F}(h|F) = \sum_{k=0} h_k F^2\log^k F
\label{re}
\ee
the scaling $F \rightarrow \lambda^2 F$ leads to the evolution equations
\be
\dot h_k = (k+1)h_{k+1},
\label{rec}
\ee
or, for $g_k = h_k/h_0$,
\be
\dot g_k = (k+1)g_{k+1} - g_1g_k.
\label{rec1}
\ee
If all $h_k$ with $k>N$ vanish, (\ref{rec}) is solved by
a polynomial $P_N(s)$ of degree $N$ for $h_0$, and
its $k$-th derivative for $h_k$. Thus
solutions  $g_k$ of (\ref{rec1}) are rational
functions of the form $g_k = \partial_s^k P_N(s)/P_N(s)$.
The scaling properties of (\ref{re}) and (\ref{rec}) can
then be employed to derive realistic multiloop $\beta$-functions
of the form
\[
\beta(g) = \sum_j b_jg^{2j+1}
\]
generalizing our one-loop result (\ref{toyF}). For this,
we substitute for $h_k$ in (\ref{re})
\be
h_0(g^2) = \frac{1}{g^2}, \nn \\
h_1(g^2) = \dot h_0 = -\frac{2\dot g}{g^3} =
-\frac{2\beta(g)}{g^3} = -2\sum_j b_jg^{2(j-1)}, \nn \\
h_2(g^2) = \dot h_1 = 2g\beta(g)h_1^\prime(g^2) =
-4\sum_{j,k} (j-1)b_jb_k g^{2(j+k-1)}, \nn \\
\ldots
\ee
If $b_0 = 0$, then
\be
h_1 = -2(b_1 + b_2g^2 + \ldots), \nn \\
h_k = -2^k  b_1^{k-1}b_2k! g^{2k} + \ldots\ \ \
{\rm for} \ k>1.
\ee

The scaling properties of the model effective actions
(\ref{toyF}), (\ref{re}) lead to
the Callan-Symanzik type equations
\be
\frac{\partial{\cal F}(t|\varphi)}{\partial\varphi} +
\sum_i \beta_i(t)
\frac{\partial{\cal F}(t|\varphi)}{\partial t_i}
+ k{\cal F}(t|\varphi) = 0
\label{CS}
\ee
where the model background field $\varphi$
in (\ref{CS}) is related to $F$ in (\ref{re}) by
$F = e^{-\varphi}$. The equation (\ref{CS}) can then
provide relations between field theoretical RG equations
and dynamical systems. In particular, we shall
now propose that issues concerning chaoticity and
integrability of the coupling flow equations can
be directly related to the construction of solutions to the
equation (\ref{CS}).

\vfill\eject
\noindent
{\bf \Large 6. Model Effective Actions And Chaos in
RG Flows}
\vskip 1.0cm

In quantum field theories, it is known that the effective
actions are usually highly complicated, multivalued
functions of the couplings. Indeed, in general
the effective actions are not ordinary function(al)s but
highly nontrivial sections of line bundles which are defined over fibrations
of the background fields $\varphi$ over the spaces
${\cal M}$ of couplings $t_i$. We shall
now propose that these complexities in their functional
form could be viewed as an indication of chaotic behaviour in the
ensuing flows of the couplings. This turns out to
be an issue that can be addressed at the level
of model effective actions
{\it i.e.} ordinary functions which model the scaling
properties of the actual field theoretical effective actions.
These model effective actions can often be constructed
by explicit integrations of the ensuing RG flow equations:

Consider the model RG equation (\ref{CS}), which we
now solve using the method of characteristics. For
this, we first need a solution to the extended system
(\ref{dynsys}),
\be
\dot t_i = \beta_i(t), \nn \\
\dot\varphi = 1
\label{dynsys2}
\ee
These solutions yield the flows $t_i(s,a_{i-1})$, with
$a_{i-1}$ as the integrations constants.
We then invert these relations, to express $s$ and the
integration constants $a_{i-1}$ as functions of the
$t_i$ and $\varphi$. The general solution to
the model RG equation (\ref{CS}) is now obtained by
introducing an {\it arbitrary} function
$f[a_{i-1}]$ of $a_{i-1}(t|\varphi)$ and setting
\be
{\cal F}(t|\varphi) = f\left[a_{i-1}(t|\varphi)\right] +
ks(t|\varphi).
\label{CSsol}
\ee
Obviously, complex behaviour such as bifurcations in the original
equations (\ref{dynsys2}) relate to
singularities in these functional inversions.

Clearly, if the equation (\ref{dynsys}) describes flow towards
a strange attractor {\it i.e. } the flow is
chaotic, the solution (\ref{CSsol}) will also lead to a
function which reflects the structure of the attractor.
Asymptotically, the model effective action (\ref{CSsol}) 
then flows towards a generalized function
or rather a (Lebesque) measure
with support on the strange attactor. Obviously this means that
we need to extend the concept of model effective actions
to include
measures with support on fractal structures. In
particular, when the flow (\ref{dynsys2}) is chaotic
the construction of its solutions can only be symbolic,
which then translates to the impossibility to construct
the model effective action (\ref{CSsol}) by quadratures.
Presumably, this could be developed into a criteria for
identifying actual field theory models where chaotic RG
flows are present.

For this, we first consider explicit constructions
of model effective actions for the flows that we have
analysed previously. We start by noting that
the last equation in (\ref{dynsys2}) is clearly
oversimplified and easily solved,
$\varphi = s + a_s$, so that $\varphi$ will actually
enters (\ref{CSsol}) only through a single integration
constant $a_s$. Moreover, it will always appear in the combination
$a_s = \varphi - s(t)$, where $s(t)$ is obtained,
together with the integration constants $a_{i-1}$, as the
functional inverse of the flows, derived from (\ref{dynsys}).
For the same reason the last term at the {\it r.h.s.} of (\ref{CSsol})
does not depend on $\varphi$ and can be written as $ks(t)$.

We first consider the flow of a single coupling,
\be
\dot t = \beta(t)
\ee
The generic solution to (\ref{CS})  is
\be
{\cal F}(t|\varphi) =
f[s(t) - \varphi] = \tilde f\left[\gamma(t)e^{-\varphi}\right],
\ee
where $f[x] = \tilde f[e^x]$ is arbitrary function of a single variable,
and $\gamma'(t)/\gamma(t) = 1/\beta(t)$, i.e.
\be
\gamma(t) = e^{s(t)} = \exp \left(\int^t \frac{dx}{\beta(x)}\right)
\ee

Similarly, if the $i$-th $\beta$-function depends on
$t_i$ only, $\beta_i(t) = \beta_i(t_i)$,
\be
{\cal F}(t|\varphi) =  \tilde f\left[\gamma_i(t_i)e^{-\varphi}\right],
\ee
where $\tilde f[x]$ is arbitrary function of its variables and
\be
\gamma_i(t_i) = \exp \left( \int^{t_i} \frac{dx}{\beta_i(x)}\right)
\ee

We then consider
\be
\dot t_i = A_{ij}t_j,
\ee
with $t$-independent matrix $A_{ij}$. According to the
Hartman-Grobman theorem \cite{HGt},
a generic flow is homeomorphic
to this near its hyperbolic fixed points. But we note that
this can also include models with limit cycles, such as
(\ref{rot}). We diagonalize $A_{ij}$ so that the system
is transformed into $\dot\xi_i = \lambda_i\xi_i$ with
\be
\xi_i = \sum_j U_{ij}t_j
\nn
\ee
\be
(UAU^{-1})_{ij} = \lambda_i\delta_{ij}
\nn
\ee
The generic solution to (\ref{CS}) then becomes
\be
{\cal F}(t|\varphi) =
f\left[\frac{\xi_i^{1/\lambda_i}}{\xi_1^{1/\lambda_1}}, \
\varphi - \frac{1}{\lambda_1}\log\xi_1\right] = \nn \\ =
\tilde f\left[\xi_i e^{-\lambda_i\varphi}\right] =
\tilde f\left[ e^{-\lambda_i\varphi} \sum_j U_{ij}t_j \right]
\ee
with arbitrary functions $f[x_i]$ and $\tilde f[x_i]$.

Finally, we consider the most general linear flow
\be
\dot t_i = A_{ij}t_j + B_i,
\ee
with $t$-independent $A_{ij}$ and $B_i$. We get
\be
{\cal F}(t|\varphi) =
\tilde f\left[  e^{-\lambda_i\varphi}\sum_j U_{ij}t_j
+ \frac{1 - e^{-\lambda_i\varphi}}{\lambda_i}\sum_j U_{ij}B_j
\right].
\ee
Note that we have defined the arguments of $\tilde f[x_i]$ so that they
make sense even if $A_{ij}$ is degenerate {\it i.e.} some of the
eigenvalues $\lambda_i$ vanish.

A particular example of the previous construction is given
by the limit cycle flow (\ref{rot}),
\be
\dot x = -\eta y, \nn \\
\dot y = \eta x
\ee
we have $x = -a \sin \eta s$, $y = a\cos\eta s$, and
the generic solution is
\be
{\cal F}(x,y|\varphi) =
f\left[\sqrt{x^2+y^2},\ \eta\varphi + \arctan(x/y)\right] =
\nn \\ =
\tilde f[x\cos\eta\varphi - y\sin\eta\varphi,\
x\sin\eta\varphi + y\cos\eta\varphi],
\ee
where $f[x,y]$ and $\tilde f[x,y]$ are arbitrary functions
of two variables.

As a further, nonlinear example we consider
the model effective action for the $\beta$-functions
(\ref{wzw}), (\ref{luki}).
For this we consider (\ref{wzw}), in the form
\be
\dot x = xy, \nn \\
\dot y = x^2
\ee
Indeed, this is the system considered in \cite{cycles1,cycles2},
with
\[
x(s) = \frac{Q}{\sin Qs} \ \ \ \ \& \ \ \ \ y(s) = Q\cot Qs
\]
The generic solution of (\ref{CS}) is
\be
{\cal F}(x,y|\varphi) =
f\left[Q,\ Q\varphi - \arccos \left(-\frac{y}{x}\right) \right]
= \tilde f\left[Q,\frac{Q\sin Q\varphi - y\cos Q\varphi}{x}, Q\right],
\ee
where $f[x,y]$ and $\tilde f[x,y]$ are arbitrary functions of
two variables, and
\[
Q(x,y) = \sqrt{x^2-y^2}
\]

It is clear that since the method of characteristics is based
on solving the system (\ref{dynsys}), in the case of chaotic flows
the method can not be explicitly implemented, even in principle.
Consequently model effective actions for chaotic flows
can not be explicitely constructed, which may also be an explanation
why concrete examples are not known in the literature.
Indeed, this is an obvious conceptual issue that explains
why it is very difficult (may be even impossible?)
to construct even simplistic model effective actions with chaotic
renormalization group flow in the first place. To exemplify
the problems encountered, we shall now consider the construction
of the model effective action for the Lorenz equations (\ref{Lore}),
by employing a perturbative expansion. A natural perturbative parameter
is $r$, which corresponds to the ratio of Rayleigh number to its
critical value in the underlying hydrodynamical model.
This suggests that we search for a perturbative construction of
the model effective action by first expanding (\ref{Lore}) around
the (chaotic) large-$r$ limit. (Recall that with the canonical
values $\sigma = 10$ and $b = 8/3$ the
model becomes chaotic only when $r > 24.74 \ .. $.)
The expansion parameter we use
is $\epsilon = 1/\sqrt{r}$,
and setting $x \to \epsilon x$, $y \to \epsilon^2 \sigma y$,
$z \to \sigma(\epsilon^2 z - 1)$ and $t \to t/\epsilon$ we get
for the Lorenz equations
\be
\dot x \ = \ y - \epsilon \sigma x \nn \\
\dot y \ = \ -xz - \epsilon y \nn \\
\dot z \ = \ xy - \epsilon b(z+\sigma)
\label{pertlor}
\ee
We then expand
\be
x \ = \ x_0 + \epsilon x_1 + \epsilon^2 x_2 + \ ... \nn \\
y \ = \ y_0 + \epsilon y_1 + \epsilon^2 y_2 + \ ... \nn \\
z \ = \ z_0 + \epsilon z_1 + \epsilon^2 z_2 + \ ...
\nn
\ee
To the leading order we get
\be
\dot x_0 \ = \ y_0 \nn \\
\dot y_0 \ = \ - x_0 z_0 \nn \\
\dot z_0 \ = \ x_0 y_0
\nn
\ee
These equations can be integrated by quadratures, in terms
of the Jacobi elliptic functions.
When we substitute the solution to the higher order equations
for $(x_n,y_n,z_n)$ ($n\geq 1$) we find at each order a set of equations
which are linear in their unknown variables. For example
at ${\cal O}(\epsilon)$ we get the linear equations
\be
\dot x_1 \ = \ y_1 - \sigma x_0 \nn \\
\dot y_1 \ = \ - x_0 z_1 - x_1 z_0 - y_0 \nn \\
\dot z_1 \ = \ x_0 y_1 + x_1 y_0 - b(z_0 + \sigma)
\nn
\ee
and so forth at higher orders in $\epsilon$.
Consequently these higher order equations
can in principle also be solved by quadratures. Thus the Lorenz equations
can be solved by quadratures, at least formally in a 
perturbation expansion to an arbitrary order in $\epsilon$.
The ensuing model effective action
can then also be constructed by employing the method of characteristics,
order by order in a perturbative expansion in $\epsilon$.
However, it turns out that this perturbative solution of the Lorenz
equations is at most asymptotic. It does not converge towards a solution
of the original Lorenz system. Instead, numerical investigations indicate
that it leads to an diverging asymptotic expansion which describes 
the chaotic solutions of the Lorenz equations
only for (relatively) small values of the flowtime $s$.

Clearly, it must be a general feature of chaotic systems
that model effective actions can not be constructed by quadratures. Not
even in principle, as this would amount to solving the original
chaotic equations. For the same reason any perturbative approach
can only lead to an asymptotic expansion, which approximates
the exact model effective action at most during a limited period of
the flow. Indeed, since a chaotic flow approaches an attractor which
is a fractal, the ensuing model effective actions must also approach
generalized functions (measures) with support on a fractal set of
points. We note that this is in a very
curious resemblance with the familiar
complex behaviour of actual effective field theory actions.
These effective actions are usually highly complicated and
multivalued function(al)s (rather section(al)s) for which any perturbative
expansion yields at most an asymptotic series.

\vfill\eject
\noindent
{\bf \Large 7. Further Developments}
\vskip 1.0cm

It was proposed already in \cite{Wor} that (at least)
cyclic behaviour of asymptotic RG flows can not be
excluded. As we have argued in the present article
even chaotic flow does not appear to conflict (most of)
the assumptions in $c$-theorems. In fact, cyclicity of the flow
seems to be quite natural in many field theory problems with
spectral flow, with an adiabatic evolution of the
energy eigenvalues.
In terms of spectral flow, a cycle can arise
whenever the energy spectrum is mapped onto itself by an
adiabatic process, but with a rearrangement: The
level $E_n$ becomes shifted into some level $E_{n-k}$.
From the point of view of an effective action, this rearrangement
of the energy levels is clearly a
cycle.

One example of nontrivial spectral flow can be developed by considering
a stringy spectrum of the form
\be
E_n = n m^2(g) + \alpha(g)
\ee
A cycle arises whenever
\be
m^2(g') = m^2(g) \ \ \ \ \ {\rm and} \ \ \ \ \
\alpha(g') = \alpha(g) - k m^2(g), \ \ k\in Z.
\ee
The intercept $\alpha(g)$ is now identified as the
dissipative Lyapunov $c$-function.
Note that in simple string models the intercept is
exactly (proportional to) the central charge of the
underlying $2d$ conformal field theory (effective space-time
dimension). Clearly, this example is very much in the spirit
of the original
introduction of the $c$-theorem in the context of conformal
field theories \cite{Zam}.

In our examples, we have inspected the $c$-theorem
in RG flows without background fields. To some
extent this can be justified, by arguing that the
background fields can be amalgamated with couplings.
However, there are also differences. For this, we
consider a hypothetical field theory model (
for example a Higgs model),
with background fields which describe a vacuum state
of the effective theory. If the parameters in the effective action
change {\it e.g.} as a consequence of coupling flow,
then the functional form of the effective action will also
change. But as the effective action changes, so does
the vacuum. Moreover, in the
case of first-order phase transitions
there are discontinuous changes in the ensuing order parameters
{\it i.e.} background fields.
Even if only second-order phase transitions occur in the course
of RG evolution and the location of vacuum is changing
continuously, its evolution is still important for
determining the effective action. The physical partition function
which emerges after accounting for the change in the vacuum
obeys RG equations somewhat different from the original
theory. For this, we consider a model effective action
\be
{\cal F}(\varphi) = \sum_{k} t_k \varphi^k
\nn
\ee
which flows as
\be
\dot{\cal F}(\varphi) = B(\varphi)   \ = \ \sum_{k} \beta_k \varphi^k.
\ee
Define a vacuum order parameter $\varphi_0$ as an extremum of
${\cal F}(\varphi)$:
\be
{\cal F}^\prime(\varphi_0) = 0.
\ee
Often we are interested in effective potential
\be
{\cal F}_0(\varphi) \equiv {\cal F}(\varphi_0 +\varphi) -
{\cal F}(\varphi_0),
\ee
which describes fluctuations around the vacuum.
Then, since
\be
\dot \varphi_0 = -B(\varphi_0)/{\cal F}^{\prime\prime}(\varphi_0)
\nn
\ee
we get
\be
\dot {\cal F}_0(\varphi) = B_0(\varphi), \nn
\ee
and
\be
B_0(\varphi) = B(\varphi_0 +\varphi) -
B(\varphi_0) -
\left[{\cal F}^\prime(\varphi_0 +\varphi) -
{\cal F}^\prime(\varphi_0)\right]
\frac{B(\varphi_0)}{{\cal F}^{\prime\prime}(\varphi_0)}
\ee
Now, even if $B(\varphi)$ is subject to a gradient flow
according to the strong $c$-theorem, there is no apparent
reason why this should be the case with $B_0(\varphi)$.

Once we account for the RG flow of the vacuum, new phenomena
become possible. For example, consider the potential
\be
V(\phi) = -\frac{m^2}{2}\phi^2 + \frac{\lambda}{4} \phi^4
\nn
\ee
The vacuum $\phi_0 = \sqrt{m^2/\lambda}$ changes smoothly
when $m^2$ and $\lambda$ are renormalized according to conventional
RG equations. Now suppose the field $\phi$ is further coupled to another
field $\chi$, for example by
\be
\chi^4\cos (\omega\phi)
\nn
\ee
Then the effective coupling $g_4$ in the $\chi^4$ vertex
is actually equal to
\be
g_4 = \cos(\omega\phi_0) = \cos (\omega \sqrt{m^2/\lambda})
\nn
\ee
and, for sufficiently large $\omega$, this $g_4$ can be
oscillating along the RG flow with ensuing changes for
$m^2$ and $\lambda$. This model could then indicate how
to construct realistic field theory models with RG
limit cycles.

An alternative is to consider
\be
V(\phi) = g_1 \cos \omega_1\phi + g_2 \cos \omega_2\phi
\ee
with $\omega_1/\omega_2$ irrational.
The potential has infinitely many irregular local minima,
which change when $g_1$ and $g_2$ are {\it smoothly} evolving
along any RG flow (which may be subject to the strong $c$-theorem).
The lowest energy minimum is a complex function of these couplings,
and the ensuing {\it  v.e.v.} $\phi_0$ can at least {\it a priori}
exhibit chaotic (irregular) behaviour.
Such irregular behaviour of $\phi_0$
can then give rise to similarly irregular behaviour of
some of the couplings.

Similar phenomena can also be realized in the context
of finite temperature field theories, with various applications
including in particular Very Early Universe and Cosmology.
In the presence of a finite temperature, the role of
flow-time $s$ is taken by the inverse temperature
$s= 1/T$. As temperature decreases, the number
of states that contribute to the partition function
decreases - the flow is contracting, irreversible.
Again, we can illustrate the phenomena with a model
partition function: We select a somewhat unconventional
set of harmonic oscillators, with potentials
\be
\tilde\alpha_i + \frac{1}{2}\omega_i^2q_i^2
\nn
\ee
The ensuing spectrum consists of oscillators
$\alpha_i + \omega_ik_i$ with integer non-negative
$k_i$, with the ground state energy
absorbed into the intercept
$\alpha_i = \tilde\alpha_i + \frac{1}{2}\omega_i$.
The partition function is
\be
Z(s) = \exp ({\cal F}(s)) =
\sum_i \frac{e^{-\alpha_i s}}{1 - e^{-\omega_i s}}
\ee
If $\alpha_i \gg \omega_i$, this can be
approximated by
\be
Z_{app}(s) = \frac{1}{s}\sum_\omega
\frac{e^{-\alpha(\omega)s}}{\omega}
\ee
In the presence of only two frequencies $\omega_1\ll\omega_2$,
with $\alpha_1\gg\alpha_2$, we then have
\be
sZ_{app}(s) = \frac{e^{-\alpha_1 s}}{\omega_1} +
\frac{e^{-\alpha_2 s}}{\omega_2},
\ee
with the first term dominating at $s\ll 1/\alpha_1$ and the
second one at $1/\alpha_1 \ll s \ll 1/\alpha_2$.
For infinitely many chaotically distributed $\omega_i$ and
$\alpha_i$, the behaviour of $Z(s)$ also exhibits
irregular (chaotic) features.
As an extreme, one can consider a somewhat peculiar
distribution of oscillators with $\omega_n \sim \sqrt{n}$ and
$\alpha_n = \alpha\log n$. Then
\be
sZ_{app}(s) = \zeta\left(\frac{1}{2} + \alpha s\right)
\ee
is the Riemann $\zeta$-function, which is suspected to
have a relation with a chaotic dynamical system.

\vfill\eject
\noindent
{\bf \Large 8. Conclusions}
\vskip 1.0cm

In conclusion, we feel that there is a clear need
to investigate the relations between RG flows
from the perspective of chaotic dynamical systems.
We have analyzed the restrictions that can be imposed
on the RG flows by employing various aspects of
$c$-theorems and local integrability conditions,
stemming from the fact that RG flows reflect the
scaling properties of effective actions. These
effective actions are in general not functions
but sections in a line bundle
over background fields and couplings. In fact, we have
argued that the involved structure of field theory
effective actions, in particular their multivaluedness
in the couplings, can be naturally understood in terms
of the impossibility to explicitely construct model
effective actions for chaotic dynamical systems. In particular,
both admit perturbative expansions which fail to converge
except in an asymptotic sense.

We have also proposed that many of the familiar
properties imposed on RG flows are not in any kind
of apparent contradiction
with the existence of chaotic behaviour. In fact,
several RG flows do reflect features such as nontrivial
vorticity which are necessary for a chaotic flow to occur.
Furthermore, we have suggested that a self-similar, chaotic RG flow in
the IR limit could actually be desirable in many applications,
{\it e.g.} to spin-glasses and neural networks.
Perhaps even more importantly to the Early Universe
Cosmology, and maybe even M-theory
with the possibility that various field theory,
string and brane models emerge at the end of its chaotic
RG trajectories. In a sense, we are then proposing that chaotic
RG orbits are as natural as the emergence of condensed
matter physics with its highly elaborate and largely self-similar
structure which emanates from the relatively
simple microscopic quantum electrodynamics.

While we feel that the idea of self-similar chaotic orbits
in field theory RG flows is
a very natural one, we can not exclude
their absence. But for this, novel no-go theorems are needed.
Such theorems should most likely be based
on conceptually new physical principles.
Whatever the necessary restrictions are,
they are bound to shed important light
to studies of hidden integrable structures of effective actions
(with complex relations between different RG flows taken into
account). They will also lead to a deeper understanding of
singularities of generalized $\tau$-functions,
and in a wide sense to
the general structure of phase transitions with applications
ranging from condensed matter to early universe and fundamental
string theories.

} 

\bigskip
\vskip 2.0cm

A.M. acknowledges the hospitality of Uppsala University.
Our work is partly supported by GSP 40.600.1.4.0025,
FNP  40.052.1.1.1112, RFBR-01-02-17488, INTAS-00561,
Volkswagen Stiftung, the Russian Grant for Support of the
Scientific Schools and by STINT Institutional Grant IG 2001-062

\vfill\eject

\end{document}